\documentclass[a4paper]{article}

\usepackage{INTERSPEECH2020}
\usepackage{multirow}
\usepackage{subfigure}
\usepackage{float}

\usepackage[font=small,skip=0pt]{caption}
\title{``This is Houston. Say again, please". The Behavox system for the Apollo-11 Fearless Steps Challenge (phase II).
}
\name{Arseniy Gorin, Daniil Kulko, Steven Grima, Alex Glasman}

\address{
Behavox Limited, Montreal, Canada}
\email{\{arseniy.gorin,daniil.kulko,steven.grima,alex.glasman\}@behavox.com}

\begin{document}

\setlength{\abovedisplayskip}{3pt}
\setlength{\belowdisplayskip}{3pt}

\maketitle
\begin{abstract}
We describe the speech activity detection (SAD), speaker diarization (SD), and automatic speech recognition (ASR) experiments conducted by the Behavox team for the Interspeech 2020 Fearless Steps Challenge (FSC-2).
A relatively small amount of labeled data, a large variety of speakers and channel distortions, specific lexicon and speaking style resulted in high error rates on the systems which involved this data. 
In addition to approximately 36 hours of annotated NASA mission recordings, the organizers provided a much larger but unlabeled 19k hour Apollo-11 corpus that we also explore for semi-supervised training of  ASR acoustic and language models, observing more than 17\% relative word error rate improvement compared to training on the FSC-2 data only.
We also compare several SAD and SD systems to approach the most difficult tracks of the challenge (track~1 for diarization and ASR), where long 30-minute audio recordings are provided for evaluation without segmentation or speaker information.
For all systems, we report substantial performance improvements compared to the FSC-2 baseline systems, and achieved a first-place ranking for SD and ASR and fourth-place for SAD in the challenge.

\end{abstract}
\noindent\textbf{Index Terms}: speech recognition, speaker diarization, speech activity detection

\section{Introduction}

The Apollo-11 corpus is not only a digitized audio recording archive of an important historical event, it is also a naturalistic dataset that demonstrates problems which are hard for state-of-the-art speech processing systems, including speech activity detection (SAD), speaker diarization (SD), automatic speech recognition (ASR), and speaker identification~\cite{Hansen2018,sangwan2013houston}.
The second Fearless Steps~\cite{fs2020is} (FS-2) extends the previous year’s inaugural challenge~\cite{Hansen2019} with streamlined diarization and automatic speech recognition tasks, which are very realistic for both researchers and engineers working with single-channel noisy recordings.

Coincidentally, we encounter similar challenges in the field of commercial speech processing for the financial industry. Trading floor recordings tend to be long, single-channel, and consist of rapid speech including short phrases without contextual information, and a specific style of speech (trading jargon, abbreviations, etc). The recordings come from a variety of commercial recording devices, and are frequently corrupted by heavy background noise and lossy compression codecs.

Our FS-2 submission and described experiments target five out of the six tracks provided by the FS-2 challenge, including SAD, SD, and ASR. For SD and ASR, both track~2 with the reference SAD, and track~1 with long audio files, are evaluated. The SAD system was ranked fourth, while SD and ASR were ranked as the top-performing systems in the FS-2 challenge\footnote{https://fearless-steps.github.io/ChallengePhase2/Final.html}.

The submitted SAD system is based on the multilayer perceptron (MLP), which has been shown to be simple and efficient~\cite{fan19auc,dwijayanti2018enhancement,ryant2013speech}. We significantly improved the performance by tuning post-processing parameters, and experimented with untranscribed data by means of ASR transcriptions of the 19k hour Apollo-11 (A11) dataset. We achieve an order of magnitude lower detection cost function (DCF) than the baseline system (0.9\% on development and 1.6\% on evaluation data versus 12.5\% and 13.6\% for the baseline).

For the SD system, we compared a state-of-the-art x-vector~\cite{snyder2018x} model trained on out-of-domain data with a conventional i-vector based system~\cite{alam2014use} trained on the FS-2 dataset only. In addition, we experimented with a combination of these representations and employed variational Bayes (VB) re-segmentation~\cite{landini2020but} to achieve 26.7\% and 27.6\% diarization error rates (DER) on the track~1 and track~2 development sets, respectively. This is a substantial improvement compared to 79.7\% and 68.6\% DER of the baseline system.

Our Kaldi based ASR~\cite{povey2011kaldi} system heavily utilizes the large and untranscribed A11 dataset by means of semi-supervised training (SST) that was shown to be useful in the past for low and moderately resourced conditions~\cite{lamel2002lightly,karafiat20172016,manohar2019semi}. In our experiments, we relied on lattice-free MMI SST~\cite{manohar2018semi} for the ``chain''~\cite{povey2016purely,hadian2018improving} acoustic models based on factorized  time-delayed neural networks (TDNN-F) proposed in~\cite{povey2018semi}. Our experiments suggest that the SST framework proposed by~\cite{manohar2018semi} is very efficient ``in the wild'', where the available large untranscribed speech data are very noisy and not segmented.
We also found that automatically transcribed data was useful for improving language models for this task. Our best performing ASR systems achieve 21.8\% and 22.4\% word error rate (WER) on tracks 2 and 1, respectively, which is almost four times better than the performance of the baseline systems (80.5\% and 83.5\%, respectively).

\section{Speech Activity Detection} \label{sec:sad}

\begin{figure}[t]
\centering
\includegraphics[width=\linewidth]{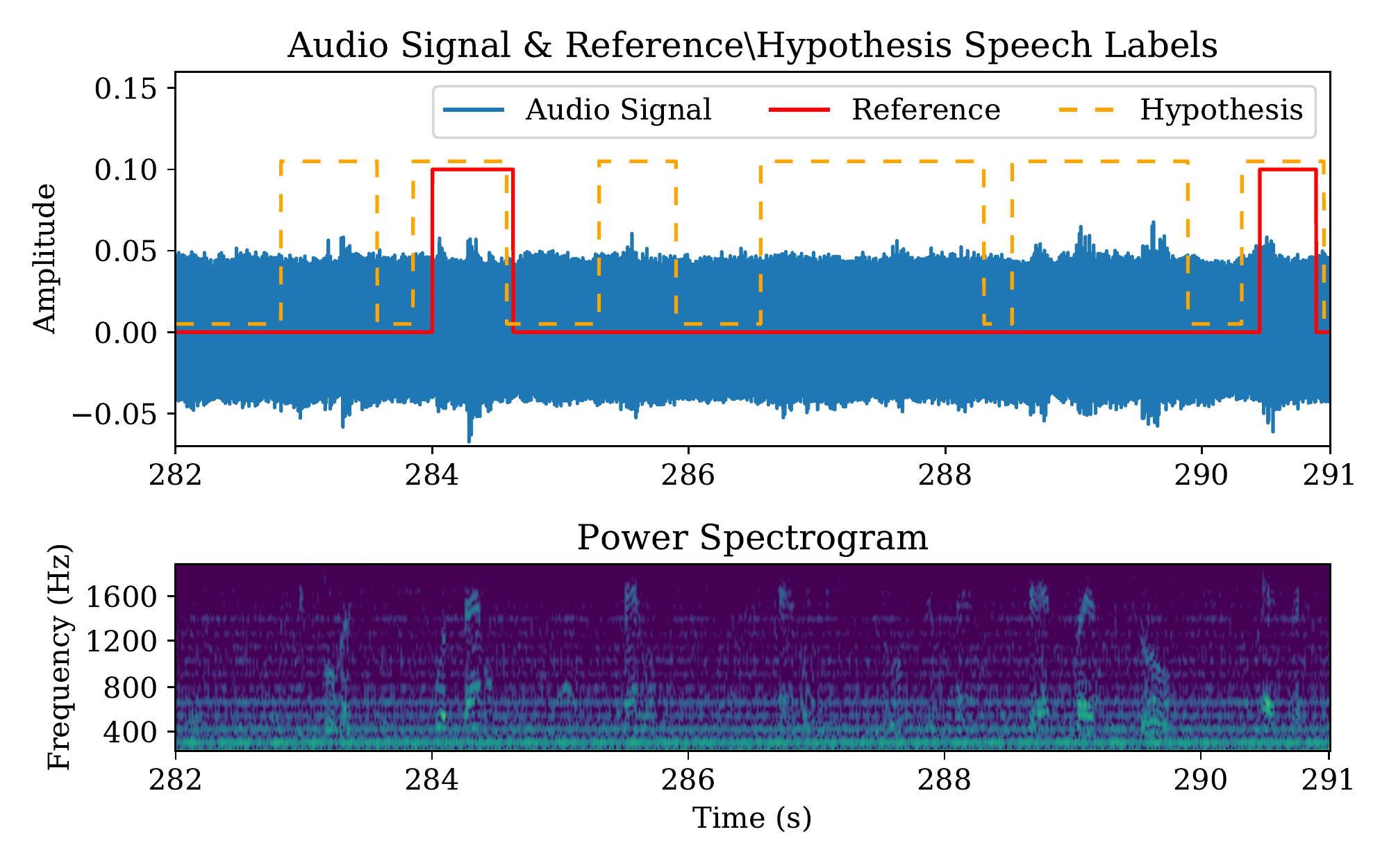}
\caption{Snippet of FS02\_dev\_020.wav [4:46-4:55] showing unlabeled speech found by the speech activity detector.}
\label{fig:vad_analysis}
\end{figure}

The SAD system is based on an MLP trained with conventional 13 MFCC features.
A context of 30 frames was used as an MLP input to produce frame-wise predictions based on a speech probability threshold $F_{thd}$. The frame-wise predictions were further grouped into segments followed by simple postprocessing, which filtered out speech segments with an average frame probability below a given threshold $S_{thd}$ or shorter than a given duration $S_{min}$ similar to~\cite{vafeiadis2019two,pham09ann}.
For all three systems summarized in Table~\ref{fig:vad_analysis}, the postprocessing parameters were selected to optimize DCF on the development data, resulting in $F_{thd}=0.02$, $S_{min}=25$ frames and $S_{thd}=0.25$.
This simple postprocessing greatly reduced false-positive speech segments, improving DCF by a large margin.

The initial SAD (system 1 in Table~\ref{tab:vad-results}) was trained on the FS-2 training data, which consisted of 18 hours of speech and 44 hours of non-speech.
It is based on a two layer MLP with 256 hidden rectified linear units (ReLUs).
To take advantage of the larger 19k hour A11 dataset, we relied on SST. First, the initial SAD was used to segment the A11 corpus. Then, the word-level (including silence and noise units) transcriptions with confidence scores were created from the ASR lattices produced by the system described later in Section~\ref{sec-asr-baseline} (system~3 in Table~\ref{tab:asr-results}).
We further selected 206 hours of non-speech and 780 hours of speech from these transcriptions and re-trained the MLP (system~2 in Table~\ref{tab:vad-results}). While using extra data did not result in a significant improvement of DCF, it allowed us to further increase the capacity of the model to three layers with 400 ReLUs resulting in our best performing SAD (system~3 in Table~\ref{tab:vad-results}).

\begin{table}[th]

\caption{Detection cost function (DCF) on development and evaluation sets of FS-2 speech activity detection challenge for raw and postprocessed classifications comparing two and thee-layer MLP trained with or without semi-supervised (A11) data.}
\label{tab:vad-results}
\centering
\begin{tabular}{ l |l | l| r r | c }
\toprule
\multirow{2}{*}{N\textsuperscript{\underline{o}}} & \multirow{2}{*}{Model size} & \multirow{2}{*}{Train data} & \multicolumn{2}{c|}{DCF Dev, \%} & DCF Eval, \% \\
&  &  & Raw  & Post. & Post. \\
\midrule
1 & 2 x 256 & FS-2 & 1.68 & 1.25 & 1.87 \\
2 & 2 x 256 & FS-2 + A11 & 3.03 & 1.23 & 2.03 \\
3 & 3 x 400 & FS-2 + A11 & 2.01 & 0.92 & 1.56 \\
\bottomrule
\end{tabular}
\vspace{-1mm}
\end{table}

Interestingly, the results indicate that the raw classification from the MLP performs better on system 1 trained with fewer data, but after applying postprocessing, systems trained with A11 data perform the best.

Assessment of SAD performance on the development set revealed that certain files had an abnormally higher false positives rate. Specifically for FS02\_dev\_020.wav the false-positive probability was more than seven times the mean of the development set. Upon further investigation, we discovered that a major factor contributing to this was missing annotations for low-level speech in heavy noise. Still the SAD presented has correctly detected such speech. One example is shown in Figure~\ref{fig:vad_analysis}.

\section{Speaker Diarization}

\begin{figure}[t]
\centering
\includegraphics[width=\linewidth]{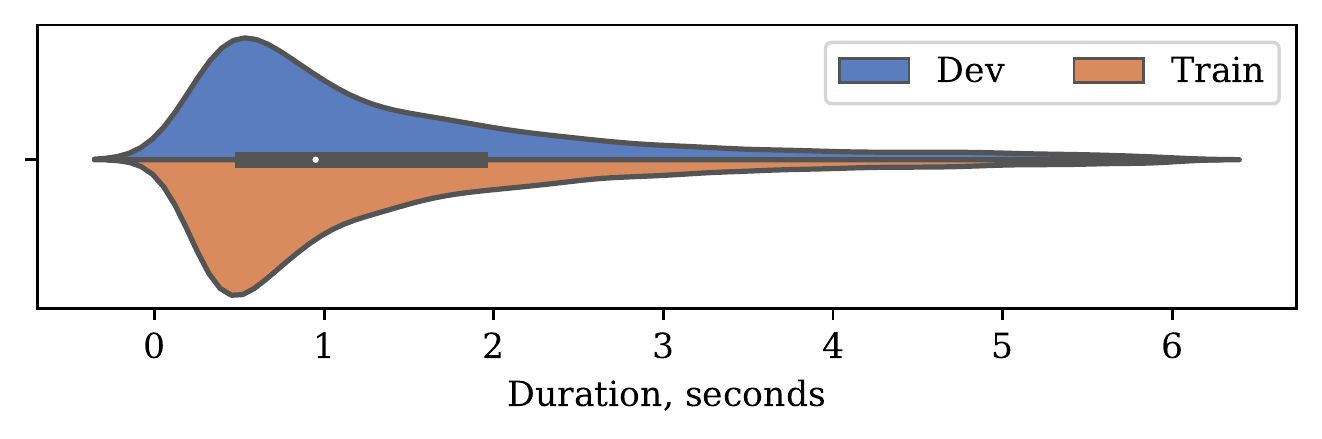}
\caption{Reference segment duration for diarization track~2.}
\label{fig:sad_duration}
\end{figure}

One challenge with the data provided for FS-2 track~2 for both SD and ASR is that it mostly consists of very short speech segments (50\% of the segments are shorter than one second) separated by long silence. Figure~\ref{fig:sad_duration} shows the distribution of reference SAD segment duration for development and training data.

\subsection{Tuning speech activity detection for  diarization}
Our diarization experiments were conducted on both track~2 (reference SAD) and track~1 (long files without SAD). Initial experiments for track~1 were done using SAD system 3 shown in Table~\ref{tab:vad-results}. However, using this SAD resulted in a large degradation on track~1 compared to track~2 (up to 1.5 times higher DER) due to many silence segments being recognized as speech. This is expected given the DCF metric definition for SAD task:
\begin{equation}
\label{eq:sad}
DCF(\theta) = 0.75 P_{FN}(\theta) + 0.25 P_{FP}(\theta)
\end{equation}
where $P_{FN}$ is false negatives duration divided by reference speech duration, $P_{FP}$ is false positive duration divided by reference silence duration, and $\theta$ denotes a threshold setting.

The imbalanced coefficients in the DCF metric encourage systems to not miss speech. However, as the FS-2 files have much more silence than speech, even small $P_{FP}$ implies a large absolute duration of false positives.
To reduce the amount of false positives for DER and the amount of insertions for WER, in this and later experiments with SD and ASR, we used the same SAD but with thresholds optimized as follows:
\begin{equation}
\label{eq:sad-inv}
DCF_{INV}(\theta) = 0.25 P_{FN}(\theta) + 0.75 P_{FP}(\theta)
\end{equation}
\subsection{Speaker diarization experiments}

We started with an i-vector based SD system described in~\cite{sell2014speaker} that used agglomerative hierarchical clustering (AHC) of probabilistic linear discriminant analysis (PLDA) scores~\cite{ioffe2006probabilistic}.
The Kaldi toolkit~\cite{povey2011kaldi} was used for modeling and implementation of the baseline similar to a publicly available Kaldi callhome diarization recipe.
In addition, we investigated neural network-based speaker embeddings (x-vectors) that have demonstrated good performance for diarization~\cite{snyder2018x,sell2018diarization}. Training an x-vector extractor generally requires a much larger amount of speaker annotated training data than is available in FS-2. We therefore used the x-vector extractor trained on out-of-domain data (NIST SRE ’04, ’05, ’06, and ’08) provided by the authors of~\cite{sell2018diarization,snyder2018x}.

In our experiments, both x-vector and i-vector extractors used 23 MFCC features. The i-vector extractor also used first and second derivatives and was trained on the FS-2 training dataset using 1024-component UBM and a 128-dimensional subspace. X-vectors were extracted from a 128-dimensional pre-softmax layer of the network described in~\cite{sell2018diarization}.
For both i/x-vectors, a global mean subtraction and the PCA whitening transform were applied to combined development and training data.

For PLDA training, the i/x-vectors were extracted from three-second chunks of speech, where speaker-dependent segments were merged within each audio file to collect sufficiently long chunks. For inference, i/x-vectors were extracted from two-second chunks with a one-second overlap. These parameters were selected to optimize DER on the development data. 
For both systems, AHC was used in each audio file for clustering PLDA scores between all pairs of vectors computed on each segment. Before estimating pairwise similarity scores, the PLDA and vectors were projected into low dimensional space using a PCA estimated per audio recording. The number of recording-dependent PCA components and the AHC stopping criteria thresholds were optimized on the development set.

The performance of i-vector and x-vector based systems with AHC is shown in rows 1 and 2 of Table~\ref{tab:dia-results}. In our experiments, both systems demonstrated similar results for track~2, but x-vectors performed better on track~1.

\begin{table}[ht]
\caption{Diarization error rate for diarization track~2 (reference segments) and track~1 (unsegmented). Comparing i-vector~(iv), x-vector~(xv) and combined~(iv+xv) systems with agglomerative hierarchical clustering (AHC), with and without variational Bayesian (VB) re-segmentation and under-clustering (UC).}
\label{tab:dia-results}
\centering
\begin{tabular}{ l|l| cc | cc }
\toprule
\multirow{2}{*}{N\textsuperscript{\underline{o}}} & \multirow{2}{*}{System} & \multicolumn{2}{|c|}{DER, track~2} & \multicolumn{2}{|c}{DER, track~1} \\
 &   & Dev & Eval & Dev & Eval \\
 \midrule
1 & iv + AHC & 30.87 & 32.10 & 34.56 & 35.98  \\
2 & xv + AHC & 30.85 & 33.20 & 30.86 & 34.78  \\
3 & iv+xv + AHC & 27.86 & 29.00 & 31.20 & 31.74 \\
\midrule
4 & + VB & 26.14 & 26.98 & 27.29 & 29.45 \\
5 & + AHC(UC) + VB & 27.65 & 26.55 & 26.73 & 28.85 \\
\bottomrule
\end{tabular}
\vspace{-1mm}
\end{table}

A fusion of x-vectors and i-vectors was shown to improve the DER in~\cite{sell2018diarization}. In our experiments, raw x-vectors and i-vectors were concatenated and the resulting 256 dimension vectors were used as an input for the same PCA and PLDA training/scoring. Such a combination (system 3 in Table~\ref{tab:dia-results}) results in a significant DER improvement on the development set of track~2 but does not improve the performance of the x-vector system for track~1.

For the two final systems, a Variational Bayesian Hidden Markov Model (VBHMM, or simply VB) re-segmentation on the vector level (described in~\cite{diez2019bayesian,landini2020but} and denoted by the authors as VBx) was used to refine the initial AHC clustering.
First, VB re-segmentation was initialized with the best performing AHC (system 3 for track~2 and system 2 for track~1 shown in Table~\ref{tab:dia-results}). The resulting system (system~4 in Table~\ref{tab:dia-results}) significantly improves DER for the corresponding AHC-based system (from 27.86\% to 26.14\% and 30.86\% to 27.29\% on the development set of tracks 2 and 1, respectively).

Next, VB re-segmentation was used with an AHC system that was tuned to produce more speakers than expected (under-clustering), while expecting VB to merge some of the fragmented speaker clusters. A similar approach was shown to be efficient in~\cite{landini2019but}, where the authors controlled AHC under-clustering by increasing the AHC stopping criterion and then relied on the ability of VB to merge some of the speaker candidates.
Instead of increasing the AHC threshold, we increased the number of recording-dependent PCA components leaving the AHC threshold unchanged.  In our preliminary experiments, this approach performed better on the development data and thus yields a different way of controlling the number of speakers. 
The performance of the diarization system using AHC with under-clustering and VB is shown in Row~5 of Table~\ref{tab:dia-results}. Overall, we observed a significant improvement on the evaluation data compared to VB on top of an optimized AHC system.

\begin{figure*}[t]
\centering
\includegraphics[width=\linewidth]{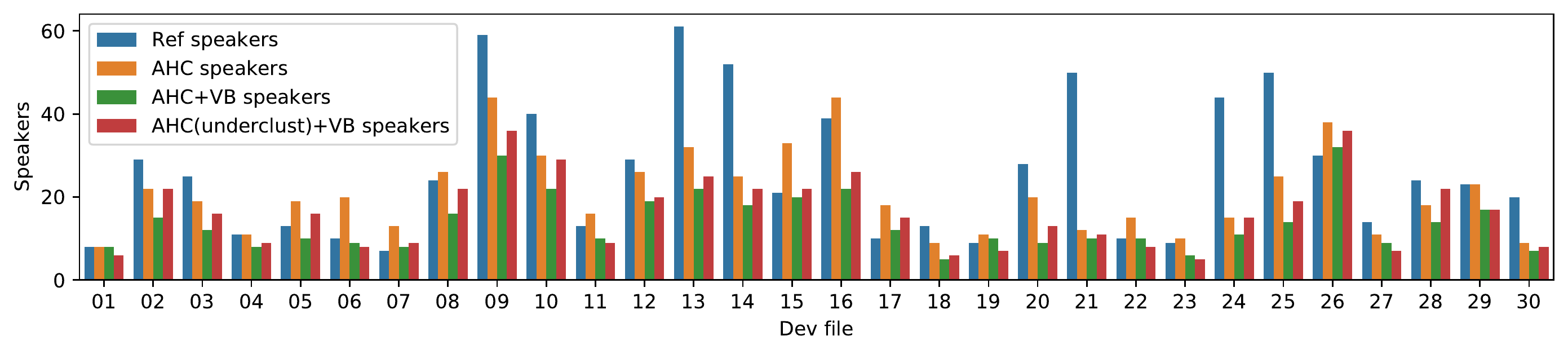}
\vspace{-5mm}
\caption{The number of reference and estimated speakers by AHC, AHC+VB and under-clustered AHC+VB systems.}
\label{fig:spk_analysis}
\vspace{-5mm}
\end{figure*}

Although the presence of the same speakers in both development and training data (potentially even evaluation data) may simplify the SD task, a large (over 50 for several files) and varying number of speakers (from seven to 61 in the development set) presents a challenge for all diarization systems. 
Figure~\ref{fig:spk_analysis} shows a per-file analysis of the reference number of speakers in the development set for track~2 along with the number of speakers predicted by AHC, AHC+VB with and without under-clustering. While errors in determining the number of speakers are lower for AHC (9.6 for AHC, 12.4 for AHC+VB and 10.8 for VB with under-clustered AHC), VB results in better DER in all our experiments.

\section{Automatic Speech Recognition}

For the ASR experiments, the Kaldi toolkit was used~\cite{povey2011kaldi}. We started by optimizing the system performance using the ground-truth SAD (track~2) and then analyzed how segmentation impacted performance on unsegmented audio (track~1).
While track~2 provides speaker labels for training and development data, no such information was released for the evaluation data. In an effort to maintain experimental consistency, we decided not to use speaker information in testing for track~2.

\subsection{Baseline automatic speech recognition system}  \label{sec-asr-baseline}

The baseline speaker adapted HMM-GMM system was trained using FS-2 training data (about 20 hours of speech). The model consists of 3k HMM states and about 100k Gaussians.
The 5k word lexicon was created using CMU dict~\cite{weide1998cmu}, and the pronunciations for the missing words were generated with sequitur toolkit~\cite{bisani2008joint}. Pronunciation and silence probabilities were also estimated from the training data alignments~\cite{chen2015pronunciation}.
Row~1 of Table~\ref{tab:asr-results} summarizes the results achieved with this system. Later, this model was used for creating training targets for neural network acoustic models (AMs).

Using the same dataset, a small TDNN-F model was trained following~\cite{povey2018semi}.
Here and in the following experiments, the frontend uses 40 MFCC features and 100-dimensional i-vector features that were shown to be efficient for introducing speaker and channel information for ASR~\cite{saon2013speaker,alam2014use}. The i-vector extractor was trained only on FS-2 training data.
Using only MFCC frontend or augmenting i-vector training data with a large subset of A11 corpus has not demonstrated good results in our preliminary experiments.
For comparison, in all experiments, a fixed context-dependent tree was used.
Given the limited amount of data, a relatively small network was trained (15 1024-dimensional layers factorized with 128-dimensional linear bottlenecks). The model was trained for 10 epochs with data augmented by speed and volume perturbation~\cite{ko2015audio}. Row~2 of Table~\ref{tab:asr-results} shows the performance of the TDNN-F baseline.

\subsection{Semi-supervised training}

We further explored the 19k hour A11 corpus for improving ASR performance. The A11 corpus also contains automatically generated transcriptions and SAD segments. However, using these annotations for training the AMs (both GMM and TDNN-F) did not improve WER in our preliminary experiments. It is likely that the provided A11 transcriptions are quite inaccurate. However, we still found it useful to add these transcriptions for language modeling (LM).

The WER achieved by including A11 transcriptions available in the dataset for LM is summarized in Row~3 of Table~\ref{tab:asr-results}. For all LM experiments we used the pocolm\footnote{https://github.com/danpovey/pocolm} toolkit that interpolates data sources at the level of data-counts to minimize the perplexity of the development set. Using A11 provided transcriptions in LM results in a significant WER improvement and an overall vocabulary expansion to 25k words. This 25k lexicon is further used in all remaining experiments.

\begin{table}[ht]
\caption{Word error rate on development and evaluation sets of FS-2 ASR  challenge track~2 (reference segmentation). Comparing models trained on the FS-2 data only, with Apollo 11 original noisy transcriptions (A11) and with semi-supervised training (SST) of the acoustic model (AM) and the language model (LM), using n-gram and recurrent neural network (RNN) LMs.}
\label{tab:asr-results}
\centering
\begin{tabular}{l|ll|ll|rr}
\toprule
\multirow{2}{*}{N\textsuperscript{\underline{o}}} & \multicolumn{2}{|c|}{AM} & \multicolumn{2}{|c|}{LM} & \multicolumn{2}{|c}{WER, \%} \\
 & Model & Data & Model & Data & Dev & Eval \\
     \midrule
1 & GMM & FS-2 & 3-gram & FS-2  & 53.8 & 55.6 \\
2 & TDNN & FS-2 & 3-gram & FS-2  & 28.6 & 31.4 \\
3 & TDNN & FS-2 & 3-gram & FS-2+A11 & 26.3 & 29.1 \\
\midrule
4 & TDNN & +SST1 & 3-gram & FS-2+A11 & 23.7 & 26.0 \\
5 & TDNN & +SST1 & 3-gram & FS-2+A11+SST1 & 23.5 & 25.8 \\
6 & TDNN & +SST1 & 4-gram & FS-2+A11+SST1 & 23.0 & 25.6 \\
7 & TDNN & +SST1 & +RNN & FS-2+A11+SST1 & 21.8 & 24.3 \\
     \midrule
8 & TDNN & +SST2 & +RNN & FS-2+A11+SST2 & 22.2 & 24.6 \\
\bottomrule
\end{tabular}
\vspace{-1mm}
\end{table}

In order to leverage A11 data for AM training, a new segmentation was generated using a lightweight SAD (system 1 in Table~\ref{tab:vad-results}). Only segments longer than two seconds and shorter than 20 seconds were selected, resulting in 980 hours of untranscribed speech segments. Next, word lattices were generated with our ASR system 3. Then, the lattices were converted to TDNN-F targets and combined with the original training data as described in~\cite{manohar2018semi}.
Supervised and transcribed samples were equally weighted and used for training a larger TDNN-F from scratch (1536-dimensional layers with 256-dimensional bottlenecks) for three epochs.
The resulting system yields 23.7\% WER on the development set (Row~4 of Table~\ref{tab:asr-results}).

1-best decoding hypotheses of A11 transcribed data were then added
for LM training.
This resulted in a small improvement for 3-gram LM (system 5 in Table~\ref{tab:asr-results}) with a significant improvement for 4-grams (system 6 in Table~\ref{tab:asr-results}).
We did not observe any improvement with 4-grams when not using semi-supervised data.
While we still found it useful to preserve the A11 original transcripts, performance was only slightly worse if discarded at this stage.

With a sufficient amount of LM data, we then trained a recurrent neural network language model (RNNLM) with letter-based features and importance sampling~\cite{xu2018neural}. See~\cite{hernandez2018ted} for a detailed description of the architecture. An efficient pruned lattice re-scoring algorithm~\cite{xu2018pruned} results in a further 1.2\% WER improvement on the development set compared to 4-gram model trained on the same data (system 7 in Table~\ref{tab:asr-results}).

Finally, we experimented with a second pass SST (system 8 in Table~\ref{tab:asr-results}), which was similar to the system 7 training methodology but used system 6 to transcribe A11 data again, this time also using a larger unsupervised set (1M segments and 1.9k hours) produced by SAD system 3. While this system did not improve track~2 WER due to an increased number of insertion errors, it did achieve the best result on track~1.

\subsection{Impact of diarization on speech recognition}

Similar to SD, ASR track~1 evaluated transcription without reference segments or speaker information. %
Table~\ref{tab:asr-results2} describes WERs on development and evaluation data of track~1, where various segmentation algorithms were compared to our two best ASR systems  (systems 7 and 8 from Table~\ref{tab:asr-results}).
The first evaluation (system~1) was performed using only segments generated by our best SAD (system~3 in Table~1).
Similarly to our SD experiments, we found that the best SAD produced too many false positives. With SAD parameters minimizing $DCF_{INV}$ (Eq.~\ref{eq:sad-inv}), the ASR performance was significantly improved (system~2 in Table~\ref{tab:asr-results2}).
Two SD systems were also compared for ASR: the simple i-vector based system (system 1 in Table~\ref{tab:dia-results}) and our best performing x-vector+VB system (system 5 in Table~\ref{tab:dia-results}). The corresponding WERs are shown in Rows 3 and 4 of Table~\ref{tab:asr-results2}.

\begin{table}[th]
\caption{Word error rate on development and evaluation sets of ASR track~1 with speech activity detector (SAD) system 3 (Table~\ref{tab:vad-results}) optimized for $DCF$ (Eq.~\ref{eq:sad}) and $DCF_{INV}$ (Eq.~\ref{eq:sad-inv}) metrics with and without speaker diarization (SD). Two best performing ASR systems are reported (systems 7 and 8 in Table~\ref{tab:asr-results}).}
\label{tab:asr-results2}
\centering
\begin{tabular}{l|l|l|rr|rr}
 \toprule
 \multirow{2}{*}{N\textsuperscript{\underline{o}}} & \multirow{2}{*}{SAD metric} & \multirow{2}{*}{SD} & \multicolumn{2}{c|}{WER, sys. 7} & \multicolumn{2}{|c}{WER, sys. 8}\\
 & & & Dev & Eval & Dev & Eval\\
  \midrule
 1 & $DCF$       & no & 24.6 & 26.1 & 23.6 & 25.0 \\
 2 & $DCF_{INV}$ & no & 23.8 & 25.2 & 22.6 & 24.1 \\
 3 & $DCF_{INV}$ & system 1 & 23.9 & 25.8 & 23.0 & 24.8 \\
 4 & $DCF_{INV}$ & system 5 & 23.1 & 24.9 & 22.4 & 24.0 \\
 \bottomrule
\end{tabular}
\vspace{-2mm}
\end{table}

The conclusions of these experiments are quite straightforward. First, similar to SD performance, better WERs are achieved with SAD optimized for the $DCF_{INV}$ metric. Second, it appears consistently better to use no diarization than a poorly performing one (compare rows 2 and 3). Finally, a good diarization system significantly improves ASR performance (compare row 2 and 4).

\section{Conclusions}
The FS-2 challenge clearly demonstrates that there is still much room for improvement of speech processing technology when working with realistic noisy recordings and a limited amount of labeled data.
Our experiments have shown that out-of-domain models can be effective (x-vector SD) and that SST is very efficient in such conditions for both acoustic and language modeling in ASR and SAD.

\section{Acknowledgements}
We would like to thank the organizers of the FS-2 challenge for coordinating this interesting and realistic task, and our colleagues Alan de Zwaan and Alex Viall for valuable suggestions.

\bibliographystyle{IEEEtran}

\bibliography{mybib_v9}

\end{document}